\documentclass[letterpaper,10pt,twoside]{article}
\usepackage{amsmath,amssymb,amsthm}
\usepackage{color}
\usepackage[dvips]{graphicx}

\newcommand{\im}{\mathrm{i}}  
\newcommand{\defeq}{:=}
\newcommand{\C}{\mathbb{C}}
\newcommand{\tens}{\otimes}   
\newcommand{\xD}{\mathcal{D}} 
\DeclareMathOperator{\id}{id}
\newcommand{\act}{\triangleright}
\newcommand{\cH}{\mathcal{H}}
\newcommand{\cS}{\mathcal{S}}
\newcommand{\cA}{\mathcal{A}}
\newcommand{\sig}{\Sigma}
\newcommand{\osig}{{\bar{\Sigma}}}

\begin{document}

\begin{titlepage}
\title{\textbf{General boundary quantum field theory: Foundations
    and probability interpretation}}
\author{Robert Oeckl\footnote{email: robert@matmor.unam.mx}\\ \\
Instituto de Matem\'aticas, UNAM Campus Morelia,\\
C.P. 58190, Morelia, Michoac\'an, Mexico}
\date{15 September 2005\\ 21 March 2006 (v2)\\ 21 August 2006 (v3)\\ with corrections of the published version (v4)}

\maketitle

\vspace{\stretch{1}}

\begin{abstract}
We elaborate on the proposed general boundary formulation as an
extension of standard quantum mechanics to arbitrary (or no) backgrounds.
Temporal transition amplitudes are generalized to amplitudes
for arbitrary spacetime regions. State spaces are associated to 
general (not necessarily spacelike) hypersurfaces.

We give a detailed foundational exposition of this
approach, including its probability interpretation and
a list of core axioms. We explain how standard quantum mechanics
arises as a special case.
We include a discussion of probability
conservation and unitarity, showing how these concepts are generalized
in the present framework. 
We formulate vacuum axioms and incorporate spacetime symmetries into
the framework. We show how the Schr\"odinger-Feynman approach is a
suitable starting point for casting quantum field theories into the
general boundary form. We discuss the role of operators.

\end{abstract}

\vspace{\stretch{1}}
\end{titlepage}

\tableofcontents
\newpage
\section{Introduction}

A key idea behind the present work is that a quantum
theory really has more structure than the standard formalism would
tell us. Notably, we suppose that transition amplitudes between
instants of time or spacelike hypersurfaces are only a special case of
what kind of amplitudes might be considered. Rather, its should be
possible to associate amplitudes with more general regions of
spacetime. At the same time, instead of a single state space, we
should have a state space associated to each boundary hypersurface of
such a region.
The single state space in the standard formalism is then only a
consequence of the restriction to spacelike hypersurfaces in connection
with a time translation symmetry.

Mathematically,
this idea may be more or less obviously motivated from the Feynman path
integral approach \cite{Fey:stnrqm} to quantum field theory: Transition
amplitudes might
be represented as path integrals. But a path integral on a region of
spacetime (should) have the property that it can be written as a
product of path integrals over parts of this region (together with
path integrals over arising boundaries). Indeed, this was the
starting point for the development of \emph{topological quantum field
theory}. This mathematical framework incorporates many of the features
of path integrals in an abstracted and idealized way.
Atiyah gave an axiomatic formulation in \cite{Ati:tqft}.
Topological quantum field theory (and its variations) have since played
an important role in quantum field theory, conformal field theory and
approaches to quantum gravity. These applications, however, have
generally not touched upon the nature of quantum mechanics itself.

The proposal we elaborate on here is of an entirely
different nature. Namely, we contend that a particular variant of the
mathematical framework of topological quantum field theory provides a
suitable context to formulate the foundations of quantum mechanics
in a generalized way. The concrete form of this formulation including
both it formal mathematical as well as its interpretational physical
aspects is what we will call the \emph{general boundary formulation}.

Since the general boundary formulation is supposed to be an
\emph{extension} rather than just a modification
of the standard formulation of quantum
mechanics, it should recover standard results in standard
situations. One might thus legitimately ask what it might be good for,
given that we are seemingly getting along very well with the
standard formulation of quantum mechanics. This is indeed true in
non-relativistic quantum mechanics as well as in quantum field theory
in Minkowski space. However, quantum field theory in curved spacetime
and, even more seriously, attempts at a quantum theory of gravity are
plagued with severe problems.

It is precisely (some of) these latter problems which motivated the
present approach. Indeed, the idea of the general boundary formulation
was proposed first
in \cite{Oe:catandclock}, motivated by the quantum mechanical
measurement problem in the background independent context of
quantum gravity. Briefly, if we want to consider a transition
amplitude in quantum gravity, we cannot interpret it naively as an
evolution between instances of time, since a classical background
time is missing. This
is the famous \emph{problem of time} in (background independent)
quantum gravity \cite{Ish:problemoftime}. If, on
the other hand, we can meaningfully assign amplitudes to regions of
spacetime having a
\emph{connected} boundary, we can avoid this problem as follows. The
state on which the amplitude is evaluated is associated with the
boundary of the region. If it is semiclassical (as we need to assume
to recover a notion of space and time) it contains spatial and temporal
information about all events on the boundary. Only if the boundary
consists of several disconnected components (as in the case of
ordinary transition amplitudes) a relation between events on different
components is lost.

Another motivation for the general boundary formulation comes from its
\emph{locality}. Amplitudes may be associated
to spacetime regions of any size. Thus, if a quantum mechanical
process is localized in spacetime, states and amplitudes associated
with a suitable region containing it
are sufficient to describe the process. In particular, we do not need
to know about physics that happens far away such as for example
the asymptotic structure of spacetime at ``infinity''.
In contrast, the standard
formulation in principle implies that we need to ``know about
everything in
the universe'', since a state contains the information about an entire
spacelike hypersurface. Of course, in non-relativistic quantum
mechanics and in quantum field theory on Minkowski space we have
suitable ways of treating isolated systems separately. However, this
is a priori not so in quantum field theory on curved spacetime. The
situation is even worse in quantum gravity due to the role of
diffeomorphisms as gauge symmetries.

It seems that there are good reasons why a general boundary
formulation should \emph{not} be feasible. On the technical side these
come from the standard quantization methods. They usually rely on a
form of the initial value problem which necessitates data on spacelike
hypersurfaces. At the same time they encode dynamics in a one-parameter
form, requiring something like a foliation. This appears to be
incompatible with the general boundary idea.
However, we contend that
this is indeed merely a technical problem that can be
overcome. Indeed, the discussion above of the motivation from path
integrals clearly points in this direction.
Note also that turning this point around yields a certain notion of
\emph{predictivity}. Clearly, the general boundary formulation is more
restrictive than the standard formulation. That is, there
will be theories that are well defined theories within standard
quantum mechanics, but do not admit an extension to the
general boundary formulation. The contention is that those theories
are not physically viable, at least not as fundamental theories.

A more fundamental reason comes from the standard interpretation
of quantum mechanics. The consistent assignment of probabilities
and their conservation seem to require a special role of time and to
single out spacelike hypersurfaces due to causality. This appears to
be in jeopardy once we try to dispense with the special role of
time.
Indeed, one is usually inclined (and this
includes quantum mechanics) to think of probabilities in terms of
something having a certain probability given that something else was
the case \emph{before}. However, a probability in general need not
have such a temporal connotation. Rather, specifying a conditional
probability that something is the case given that something else is
the case can be perfectly sensible without the presence of a definite
temporal relation between the facts in question. This is indeed the
principle on which the probability interpretation proposed in this
work rests. The standard probability interpretation arises merely as a
special case of this.

Apart from its mathematical motivation there is
is also a good physical reason to believe that a general boundary
interpretation \emph{should} exist \cite{Oe:catandclock}. This comes
from quantum field theory in the guise of \emph{crossing
symmetry}. When deriving the S-matrix in perturbative quantum field
one finds that the resulting amplitude puts the incoming and outgoing
particles practically on the same footing. This suggests that it is
sensible to think about them as being part of the same single state
space associated with the initial and final hypersurface
together. What is more, it suggests that the S-matrix may be derived
as the asymptotic limit of the amplitude associated with a spacetime
region with \emph{connected} boundary. A possible context for this is
discussed in the companion paper \cite{Oe:KGtl}.

A first step to elaborate on the idea
of the general boundary formulation was taken in \cite{Oe:boundary},
with the proposal of a list of core axioms. These were formulated in
such a way as to be applicable to a variety of background structures,
including the possibility of no (metric) background at all.
At the same time, a
tentative analysis of its application to non-relativistic quantum
mechanics, quantum field theory and 3-dimensional quantum gravity was
performed.
Unsurprisingly, the general boundary formulation much
more naturally applies to quantum field theory rather than to
non-relativistic quantum mechanics. This is because it is based on
spacetime notions, while in non-relativistic quantum mechanics the
notion of space is secondary to that of time. This motivated the
choice of the name \emph{general boundary quantum field theory} in the
title. Unfortunately, this also means that the in other circumstances
good idea of ``trying out'' non-relativistic theories with
finitely many degrees of freedom first is not particularly useful
here.

An important step in demonstrating the feasibility of the general
boundary formulation was performed in \cite{Oe:timelike}: It was shown
that states on \emph{timelike} hypersurfaces in quantum field theory
are sensible. The example discussed was that of timelike hyperplanes
in the Klein-Gordon theory. This example is considerably extended in
the companion paper \cite{Oe:KGtl}, where a further type of timelike
hypersurfaces is considered (the hypercylinder). In particular, this
provides the first example of amplitudes associated to regions with
connected boundaries. All properties of the framework are tested there,
including composition of amplitudes, the vacuum state, particles and
the probability interpretation.

Based on these experiences,
we present here a considerably deeper and more extensive
treatment of the general boundary formulation, turning it from an idea
into a definite framework. This includes, firstly, a
refined and extended list of axioms (Section~\ref{sec:axioms}). The
main additional structure
compared to the treatment in \cite{Oe:boundary} is an inner product on
state spaces. This is instrumental for what we consider the most
important part of the present work, namely the probability
interpretation (Section~\ref{sec:prob}). (Section~\ref{sec:recover}
covers the recovery of the standard formulation.)
Thereby, we hope to provide a
physically fully satisfactory interpretation of general boundaries,
which thusfar has been missing.

Further subjects covered are a proposal for an axiomatic
characterization of a vacuum state (Section~\ref{sec:vacuum}), a
discussion of various background structures and the incorporation of their
spacetime symmetries in axiomatic form (Section~\ref{sec:bgsym}).
We then proceed to elaborate on how the Feynman path integral together
with the Schr\"odinger representation may provide a viable approach to
cast quantum field theories into general boundary form
(Section~\ref{sec:schrfeyn}). We also discuss to exactly which types of
spacetime regions it is (or should be) permissible to associate
amplitudes (Sections~\ref{sec:size} and \ref{sec:corners}). Finally,
we make some remarks on the role of operators in the formalism
(Section~\ref{sec:ops}). We end with some conclusions
(Section~\ref{sec:concl}).

\section{Core axioms}
\label{sec:axioms}

The core idea of the general boundary formulation might be summarized
very briefly as follows: We may think of quantum mechanical processes
as taking place in regions of spacetime with the data to describe them
associated to the regions' boundaries. To make this precise we
formulate a list of axioms, referred to in the following as the
\emph{core axioms}. These extend and refine the axioms
suggested in \cite{Oe:boundary}. We preserve the numbering from that
paper denoting additional axioms with a ``b''.
The main addition consists of inner
product structures. As one might suspect,
these are instrumental in a probability
interpretation which is the subject of Section~\ref{sec:prob}.

The spacetime objects to appear in the axioms are of two kinds:
\emph{regions} $M$ and \emph{hypersurfaces} $\sig$. What these are
exactly depends
on the background structure of the theory in question. We will discuss
this in Section~\ref{sec:bgsym}.
If we are interested in standard quantum field theory, spacetime is
Minkowski space. The regions $M$ are then 4-dimensional submanifolds
of Minkowski space and the boundaries $\sig$ are oriented
hypersurfaces (closed 3-dimensional submanifolds) in Minkowski
space. Orientation here means that we choose a ``side'' of the
hypersurface. Given a region $M$, its boundary is naturally
oriented.\footnote{To be more explicit, any 4-dimensional submanifold of
  Minkowski space inherits a globally chosen orientation of Minkowski
  space. It is this orientation that induces the orientation of the
  boundary. If we are in a situation of not having a globally oriented
spacetime background, we need to explicitly specify an orientation of
the region to induce an orientation on its boundary.}
To be specific, we think of this orientation as choosing the
``outer side'' of the boundary. Furthermore, not all 4-dimensional
submanifolds are admissible as regions. However, this restriction is of
secondary importance for the moment and we postpone its discussion
to Section~\ref{sec:size}.

Given an oriented hypersurface $\sig$
we denote the same hypersurface with opposite orientation by $\osig$,
i.e., using an over-bar. For brevity, we use the term hypersurface to mean
oriented hypersurface.

\begin{itemize}
\item[(T1)]
 Associated to each hypersurface $\sig$ is a complex vector
 space  $\cH_\sig$, called the \emph{state space} of $\sig$.
\item[(T1b)]
 Associated to each hypersurface $\sig$ is an antilinear
 map $\iota_\sig:\cH_{\sig}\to\cH_{\osig}$. This map is an involution
 in the sense that $\iota_\osig\circ\iota_\sig=\id_{\sig}$ is the
 identity on $\cH_\sig$.
\item[(T2)]
 Suppose the hypersurface $\sig$ is a disjoint union of
 hypersurfaces, $\sig=\sig_1\cup\cdots\cup\sig_n$. Then, the state
 space of $\sig$ decomposes into a tensor product of state spaces,
 $\cH_\sig=\cH_{\sig_1}\tens\cdots\tens\cH_{\sig_n}$.
\item[(T2b)]
 The involution $\iota$ is compatible with the above
 decomposition. That is, under the assumption of (T2),
 $\iota_\sig=\iota_{\sig_1}\tens\cdots\tens\iota_{\sig_n}$.\footnote{Here
 as in the following we commit a slight abuse of notation by using the
 tensor product symbol even when considering maps that are not
 $\C$-linear, but rather $\C$-antilinear in one or more
 components. However, the meaning should always be clear from the
 context.} 
\item[(T3)]
 For any hypersurface $\sig$, there is a non-degenerate bilinear pairing
 $(\cdot,\cdot)_\sig:\cH_\osig\tens\cH_\sig\to\C$. This pairing is
 symmetric in the sense that $(a,b)_\sig=(b,a)_\osig$.
 Furthermore, the pairing is such that it induces a positive
 definite hermitian inner product $\langle\cdot,\cdot\rangle_\sig 
 \defeq(\iota_\sig(\cdot),\cdot)_\sig$ on
 $\cH_\sig$ and turns $\cH_\sig$ into a Hilbert space.
\item[(T3b)]
 The bilinear form of (T3) is compatible with the decomposition of
 (T2). Thus, for a hypersurface $\sig$ decomposing into disconnected
 hypersurfaces $\sig_1$ and $\sig_2$ we have
 $(a_1\tens a_2,b_1\tens b_2)_\sig=(a_1,b_1)_{\sig_1}
 (a_2,b_2)_{\sig_2}$.
\item[(T4)]
 Associated with each region $M$ is a linear map from the
 state space of its  boundary $\sig$ to the complex numbers,
 $\rho_M:\cH_\sig\to\C$. This is called the \emph{amplitude}
 map. 
\item[(T4b)]
 Suppose $M$ is a region with boundary $\sig$, consisting of two
 disconnected components, $\sig=\sig_1\cup\sig_2$. Suppose the
 amplitude  map $\rho_M:\cH_{\sig_1}\tens\cH_{\sig_2}\to\C$
 gives rise to an isomorphism of vector spaces
 $\tilde{\rho}_M:\cH_{\sig_1}\to\cH_{\osig_2}$.
 Then we require $\tilde{\rho}_M$ to preserve the inner product, i.e.,
 be \emph{unitary}.
\item[(T5)]
 Let $M_1$ and $M_2$ be two regions such that the union
 $M_1 \cup M_2$ is again a region and the intersection is a
 hypersurface $\sig$. Suppose that $M_1$ has a boundary with
 disconnected components $\sig_1\cup\sig$ and $M_2$ has a boundary with
 disconnected components $\osig\cup\sig_2$. Suppose amplitude maps on
 $M_1$,  $M_2$ and $M_1\cup M_2$ induce maps
 $\tilde{\rho}_{M_1}:\cH_{\sig_1}\to\cH_{\osig}$,
 $\tilde{\rho}_{M_2}:\cH_{\osig}\to\cH_{\osig_2}$ and
 $\tilde{\rho}_{M_1\cup M_2}:\cH_{\sig_1}\to\cH_{\osig_2}$.
 We require then $\tilde{\rho}_{M_1\cup M_2}
 =\tilde{\rho}_{M_2}\circ\tilde{\rho}_{M_1}$.
\end{itemize}

Before coming to the physics let us make some mathematical remarks. In
contrast to \cite{Oe:boundary} we are here much more careful about the
expected infinite dimensional nature of the state spaces. This is the
reason for example for the reformulation of axiom (T3). In
\cite{Oe:boundary} it simply stated that the state space of an
oppositely oriented hypersurface be identified with the dual of the
state space of the original hypersurface. Thus, for consistency the
bidual space must be identified with the original one. For an infinite
dimensional space this is not the case for the naively
defined dual. Here, we use the involution $\iota$ and require that a
Hilbert space structure is induced on the state spaces. Note that
this implies that $\cH_\osig$ is the Hilbert space dual of $\cH_\sig$
and consequently, the bidual is canonically isomorphic to the
original space, as required.

The tensor product in (T2) is to be understood to be the tensor
product of Hilbert spaces and not merely the algebraic one. To make
this more clear, (T3) might have been moved before (T2), but we
decided to conserve the numbering of \cite{Oe:boundary}.

Note that for the amplitude map of axiom (T4) we may ``dualize''
boundaries (as stated explicitly in the version of
\cite{Oe:boundary}). This means that if 
the boundary $\sig$ of a region $M$ decomposes into disconnected
components $\sig_1\cup\cdots\cup\sig_n$ the amplitude map $\rho_M$
gives rise to a map
$\tilde{\rho}_M:\cH_{\sig_1}\tens\cdots\tens\cH_{\sig_k}
\to \cH_{\osig_{k+1}}\tens\cdots\tens\cH_{\osig_n}$. This is simply
obtained by dualizing the tensor components
$\cH_{\sig_{k+1}}$, \dots, $\cH_{\sig_n}$. Actually, it is not
guaranteed that $\tilde{\rho}_M$ exists, the obstruction being that
the image of a state might not be normalizable. Such an induced map
(if it exists) is used in axioms (T4b) and (T5). Note that we could
formulate (T5) also with the original amplitude maps by inserting in
the pair of Hilbert spaces for the common boundary a Hilbert basis
times its dual.

We now turn to the physical meaning of the axioms. The state spaces of
axiom (T1) are supposed to represent in some way spaces of physical
situations. In contrast to the standard formalism, a state
is not in general supposed to encode ``the situation of the whole
world''. Rather, (as we shall see in more detail in the probability
interpretation) it may be thought of as encoding some ``knowledge''
about a physical situation or more concretely, an
experiment. Furthermore, the localization in spacetime of the
hypersurface to which it is associated has the connotation of
localization of knowledge about a process or measurement. Another
possible connotation is that of information (encoded in states)
``flowing'' through the hypersurface.

The axiom (T1b) serves to enable us to identify a state on a
hypersurface with the state on the ``other side'' of that hypersurface
that has the same physical meaning. Axiom (T2) tells us that the
physical situations (or information) associated to disconnected
hypersurfaces is a priori ``independent''. (Recall that the Hilbert
space of a system of two independent components in the standard
formulation is the tensor product of the individual Hilbert spaces.)
Axiom (T3) establishes the
inner product and thus lets us decide when states (e.g., experimental
circumstances) are mutually exclusive.

Axiom (T4) postulates an amplitude map. The name ``amplitude'' is chosen
to reflect the fact that this amplitude map serves to generalize the
concept of transition amplitude in the standard formulation. An
amplitude here is associated to a region of spacetime. This
generalizes the time interval determining a transition amplitude. The
idea is that the process we are trying to describe takes place in this
spacetime region. At the same time the knowledge or information we use
in its description resides on the (state spaces of the) boundary.

Axiom (T4b) says roughly the following:
If we take a state on $\sig_1$, evolve it along $M$ to $\osig_2$,
conjugate it via $\iota$ to $\sig_2$, evolve it back along $M$ to
$\osig_1$, conjugate again via $\iota$ to $\sig_1$, then we get back
the original state.
As we shall see, this axiom is responsible for a notion of
probability conservation, generalizing the corresponding notion of
temporal probability conservation in the standard
formulation. 
Axiom (T5) may be 
described as follows: Given a state on $\sig_1$, evolving it first
along $M_1$ to
$\osig$ and then along $M_2$ to $\osig_2$ yields the same result as
evolving it directly from $\sig_1$ to $\osig_2$ along $M_1\cup M_2$.
This axiom describes the composition of processes and generalizes the
composition of time evolutions of the standard formulation.

\section{Recovering the standard formulation}
\label{sec:recover}

\begin{figure}
\begin{center}
\input{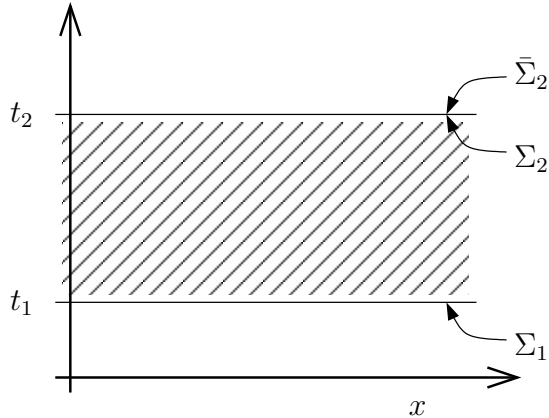}
\end{center}
\caption{The standard setup of a pair of equal-time hyperplanes
  enclosing a time interval.}
\label{fig:standard}
\end{figure}

The explanation of the physical meaning of the axioms so far has been
rather vague. We proceed in the following to make it concrete. The
first step in this is to show how exactly the standard formulation is
recovered. This clarifies, in particular, in which sense the proposed
formulation is an extension of the standard one, rather than a
modification of it.

Suppose we are interested in a quantum process, which in the standard
formalism is described through a transition amplitude from a time $t_1$
to a time $t_2$. The
spacetime region $M$ associated with the process is the the time
interval $[t_1,t_2]$ times all of space. The boundary $\partial M$ of $M$
consists of two disconnected components $\sig_1$ and $\osig_2$, which are
equal-time hyperplanes at $t_1$ and $t_2$ respectively. Note that they
have opposite orientation. $\sig_1$ is oriented towards the past and
$\osig_2$ towards the future. This is illustrated in
Figure~\ref{fig:standard}.
Using axiom (T2) the total state space $\cH_{\partial M}$ (postulated
by (T1)) decomposes into the tensor product
$\cH_{\sig_1}\tens\cH_{\osig_2}$ of state
spaces associated with these hyperplanes. Thus, a state in
$\cH_{\partial M}$ is a linear combination of states obtained as
tensor products of states in $\cH_{\sig_1}$ and $\cH_{\osig_2}$.

Now consider a state $\psi_{\sig_1}$ on $\sig_1$ and a state
$\eta_{\osig_2}$ on $\osig_2$. We will make use of axiom (T1b) to
convert $\eta_{\osig_2}$ to a state $\eta_{\sig_2}\defeq
\iota_{\osig_2}(\eta_{\osig_2})$ on the same hyperplane, but with the
opposite orientation (i.e., oriented as $\sig_1$).
Consider the amplitude $\rho_M:\cH_{\sig_1}\tens\cH_{\osig_2}\to\C$
postulated by axiom (T4). It induces a linear map
$\tilde{\rho}_M:\cH_{\sig_1}\to \cH_{\sig_2}$ in the manner described
above. We may thus rewrite the amplitude as
\begin{multline*}
 \rho_M(\psi_{\sig_1}\tens\eta_{\osig_2})
 =(\eta_{\osig_2},
  \tilde{\rho}_M(\psi_{\sig_1}))_{\sig_2} \\
 =(\iota_{\sig_2}(\eta_{\sig_2}),
  \tilde{\rho}_M(\psi_{\sig_1}))_{\sig_2}
 =\langle\eta_{\sig_2},
  \tilde{\rho}_M(\psi_{\sig_1})\rangle_{\sig_2},
\end{multline*}
where $(\cdot,\cdot)_{\sig_2}$ is the bilinear pairing of axiom (T3) and
$\langle\cdot,\cdot\rangle_{\sig_2}$ is the induced inner product.

The final expression represents the
transition amplitude from a state $\psi_{\sig_1}$ at time $t_1$ to a
state $\eta_{\osig_2}$ at time $t_2$. What appears to be different
from the standard formulation is that the two states live in different
spaces (apart from the fact that one would be a ket-state and the
other a bra-state). However, as we shall see later
(Section~\ref{sec:sym}), we may use time-translation
symmetry to identify all state spaces associated to (past-oriented
say) equal-time hypersurfaces. This is then \emph{the} state space
$\cH$ of the standard formalism.
Consequently, the
linear map $\tilde{\rho}_M$ is then an operator on $\cH$, namely the
time-evolution operator. Given that $\tilde{\rho}_M$ is invertible (as
it should be,
see the discussion in Section~\ref{sec:invevol}) axiom (T4b) ensures
its unitarity. Note that axiom
(T5) ensures in this context the composition property of
time-evolutions. Namely, evolving from time $t_1$ to time $t_2$ and
then from time $t_2$ to time $t_3$ is the same as evolving directly
from time $t_1$ to time $t_3$.

Thus, we have seen how to recover standard transition amplitudes and
time-evolution from the present formalism. Indeed, we could
restrict the allowed hypersurfaces to equal-time hyperplanes and the
allowed regions to time intervals times all of space. Then, the
proposed formulation would be essentially equivalent to the standard
one. Of course, the whole point is that we propose to admit more
general hypersurfaces and more general regions. 

Starting from a theory in the standard formulation the challenge is
two-fold. Firstly, we need to show that the extended structures (state
spaces, amplitudes etc.) exist, are coherent (satisfy the axioms) and
reduce to the standard ones as described above. This is obviously
non-trivial, i.e., a given theory may or may not admit such an
extension. We have argued elsewhere \cite{Oe:catandclock} that
crossing symmetry of the S-matrix (as manifest for example in the LSZ
reduction scheme) is a very strong hint that generic quantum field
theories do admit such an extension.

Secondly, we need to give a \emph{physical} interpretation to
these new structures. A key element of the physical interpretation in
the standard formalism is the possibility to interpret the modulus
square of the transition amplitude as a probability. In the context
above its is clear that
$|\rho_M(\psi_{\sig_1}\tens\eta_{\osig_2})|^2$ denotes the
probability of
observing the state $\eta_{\osig_2}$ given that the state
$\psi_{\sig_1}$ was prepared. Indeed, the modulus square of the
amplitude function generally plays the role of an (unnormalized)
probability. The details of the probability interpretation in the
general boundary formulation, constituting perhaps the most
significant aspect of the present work, are discussed in the following
section.

\section{Probability interpretation}
\label{sec:prob}

\subsection{Examples from the standard formulation}

To discuss the probability interpretation we start with a review of it
in the standard formulation. Let $\psi\in\cH_1$ be the (normalized)
ket-state
of a quantum system at time $t_1$, $\eta\in\cH_{\bar{2}}$ a (normalized)
bra-state at time $t_2$.\footnote{$\cH_{\bar{2}}$ indicates
a space of bra-states, i.e., the Hilbert dual of the space
$\cH_2$ of ket-states. Usually of course one considers only
  one state space, i.e., $\cH_1$ and $\cH_2$ are canonically
  identified. We distinguish them here formally to aid the later
  comparison with the general boundary formulation.}
The associated transition
amplitude $A$ is given by $A=\langle \eta|U|\psi\rangle$, where
$U:\cH_1\to\cH_2$ is the time-evolution operator of the system,
evolving from time
$t_1$ to time $t_2$. The associated probability $P$ is the modulus
square of $A$, i.e., $P=|A|^2$. What is the physical meaning of $P$?
The simplest interpretation of this quantity is as expressing the
probability of finding the state $\eta$ at time $t_2$ given that the
state $\psi$ was prepared at time $t_1$. Thus, we are dealing with a
conditional probability. To make this more explicit let us write it as
$P(\eta|\psi)$ (read: the probability of $\eta$ conditional on
$\psi$). An important ingredient of this interpretation is that the
cumulative probability of all exclusive alternatives is $1$. The
meaning of the latter is specified with the help of the inner
product. Thus, let $\{\eta_i\}_{i\in I}$ be an orthonormal basis of
$\cH_{\bar{2}}$,
representing a complete set of mutually exclusive measurement
outcomes. Then, $\sum_{i\in I} P(\eta_i|\psi)=\sum_{i\in I}\langle
\eta_i|U|\psi\rangle=1$.

This interpretation might be extended in obvious ways. Suppose for
example that we know a priori that only certain measurement outcomes
might occur. (We might select a suitable subset of performed
measurements.)
A way to formalize this is to say that the possible
measurement outcomes lie in a (closed) subspace $\cS_{\bar{2}}$ of
$\cH_{\bar{2}}$. Suppose $\{\eta_i\}_{i\in J}$ is an orthonormal basis
of $\cS_{\bar{2}}$. We are now interested in the probability of a
given outcome specified by a state $\eta_k$ conditional both on the
prepared state
being $\psi$ and knowing that the outcome must lie in
$\cS_{\bar{2}}$. Denote this conditional probability by
$P(\eta_k|\psi,\cS_{\bar{2}})$. To obtain it we
must divide the conditional probability $P(\eta_k|\psi)$ by the
probability $P(\cS_{\bar{2}}|\psi)$ that the outcome of the
measurement lies in
$\cS_{\bar{2}}$ given the prepared state is $\psi$. This is simply
$P(\cS_{\bar{2}}|\psi)=\sum_{i\in J}P(\eta_i|\psi)=\sum_{i\in J}|\langle
\eta_i|U|\psi\rangle|^2$. Supposing the result is not zero (which would
imply the impossibility of obtaining any measurement outcome in
$\cS_{\bar{2}}$ and thus the meaninglessness of the quantity
$P(\eta_k|\psi,\cS_{\bar{2}})$), 
\[
 P(\eta_k|\psi,\cS_{\bar{2}})=\frac{P(\eta_k|\psi)}{P(\cS_{\bar{2}}|\psi)}
 =\frac{|\langle \eta_k|U|\psi\rangle|^2}
 {\sum_{i\in J}|\langle \eta_i|U|\psi\rangle|^2} .
\]

We can further modify this example by testing not against a single
state, but a closed subspace $\cA_{\bar{2}}\subseteq\cS_{\bar{2}}$,
denoting the associated
conditional probability by $P(\cA_{\bar{2}}|\psi,\cS_{\bar{2}})$. This
is obviously
the sum of conditional probabilities $P(\eta_k|\psi,\cS_{\bar{2}})$ for an
orthonormal basis $\{\eta_i\}_{i\in K}$ of $\cA_{\bar{2}}$ (we suppose here
that the orthonormal basis of $\cS_{\bar{2}}$ is chosen such that it restricts
to one of $\cA_{\bar{2}}$). That is,
\[
P(\cA_{\bar{2}}|\psi,\cS_{\bar{2}})=\frac{\sum_{i\in K}|\langle
 \eta_i|U|\psi\rangle|^2}
 {\sum_{i\in J}|\langle \eta_i|U|\psi\rangle|^2} .
\]

A conceptually different extension is the following. Suppose
$\{\psi_i\}_{i\in I}$ is an orthonormal basis of $\cH_1$. Then, the
quantity
$P(\psi_k|\eta)=|\langle \eta|U|\psi_k\rangle|^2$ describes the
conditional probability of the prepared state having been $\psi_k$
given that $\eta$ was measured.
This may be understood in the following sense. Suppose somebody
prepared a large sample of measurements with random choices of initial
states $\psi_i$.\footnote{Here as elsewhere in the
  elementary discussion of probabilities we may assume for simplicity
  that state spaces are finite dimensional. This avoids difficulties
  of the infinite dimensional case which might require the
  introduction of probability densities etc.}
We then perform measurements as to whether the final
state is $\eta$ or not (the latter meaning that it is orthogonal to
$\eta$). The probability distribution of the the
initial states $\psi_k$ in the sample of measurements resulting in
$\eta$ is then given by $P(\psi_k|\eta)$.

These examples are supposed to illustrate two points. Firstly,
the modulus square of a transition amplitude might be interpreted as a
conditional probability in various different ways.
Secondly, the roles of
different parts of a measurement process in respect to which is
considered conditional one which other one are not fixed.
In particular, the interpretation is not restricted to ``final state
conditional on initial state''.

\subsection{Probabilities in the general boundary formulation}

These considerations together with the general philosophy of the
general boundary context lead us to the following formulation of the
probability interpretation. Let $\cH$ be the the generalized state
space describing a given physical system or measurement setup (i.e.,
it is the state space associated with the boundary of the spacetime
region where we consider the process to take place).
We
suppose that a certain knowledge about the process amounts to the
specification of a closed subspace $\cS\subset\cH$. That is, we assume
that we know the state describing the measurement process to be part
of that subspace. Say we are now interested in evaluating whether the
measurement outcome corresponds to a closed subspace
$\cA\subseteq\cS$. That is, we are interested in the conditional
probability $P(\cA|\cS)$ of the measurement process being described by
$\cA$ given that it is described by $\cS$. Let $\{\xi_i\}_{i\in I}$ be
an orthonormal basis of $\cS$ which reduces to an orthonormal basis
$\{\xi_i\}_{i\in J\subseteq I}$
of $\cA$. Then,
\[
 P(\cA|\cS)=\frac{\sum_{i\in J}|\rho(\xi_i)|^2}
 {\sum_{i\in I}|\rho(\xi_i)|^2} .
\]
By construction, $0\le P(\cA|\cS)\le 1$. (Again it is assumed that the
denominator is non-zero. Otherwise, the conditional probability would
be physically meaningless.) One might be tempted to interpret the
numerator and the denominator separately as probabilities. However,
that does not appear to be meaningful in general. As a special case,
if $\cA$ has dimension one, being spanned by one normalized vector
$\xi$ we also write $P(\cA|\cS)=P(\xi|\cS)$.

Let us see how the above examples of the probability
interpretation in the standard formulation are recovered. Firstly, we
have to suppose that the state space factors into a tensor product of
two state spaces, $\cH=\cH_1\tens\cH_{\bar{2}}$.
Select a state $\psi\in\cH_1$
and set $\cS_\psi\defeq\{\xi\in\cH|\exists \eta\in\cH_{\bar{2}}:
\xi=\psi\tens\eta\}\subset\cH$.
Let us denote by $\{\psi\tens\eta_i\}_{i\in I}$ an orthonormal basis of
$\cS_\psi$. Then, the probability of ``observing
$\eta\in\cH_{\bar{2}}$'' subject
to the ``preparation of $\psi\in\cH_1$'' turns out as
\[
 P(\psi\tens\eta|\cS_\psi)=\frac{|\rho(\psi\tens\eta)|^2}
 {\sum_{i\in I}|\rho(\psi\tens\eta_i)|^2} .
\]
Comparing the notation to the standard formalism, i.e., recognizing
$\rho(\psi\tens\eta)=\langle \eta|U|\psi\rangle$ shows that we recover
the standard result $P(\eta|\psi)$, up to a normalization factor,
depending as it seems on $\psi$.

Similarly, the second example is recovered by setting
$\cS_{(\psi,\cS_{\bar{2}})}\defeq\{\xi\in\cH|\exists \eta\in\cS_{\bar{2}}:
\xi=\psi\tens\eta\}\subset\cH$. Taking an orthonormal basis
$\{\psi\tens\eta_i\}_{i\in J}$, we get agreement of
$P(\psi\tens\eta_k|\cS_{(\psi,\cS_{\bar{2}})})$ with
$P(\eta_k|\psi,\cS_{\bar{2}})$
(with correct normalization). The modified example is recovered with
$\cA_{(\psi,\cA_{\bar{2}})}\defeq\{\xi\in\cH|\exists \eta\in\cA_{\bar{2}}:
\xi=\psi\tens\eta\}\subset\cH$ via
$P(\cA_{\bar{2}}|\psi,\cS_{\bar{2}})
=P(\cA_{(\psi,\cA_{\bar{2}})}|\cS_{(\psi,\cS_{\bar{2}})})$.
For the third example set
$\cS_\eta\defeq\{\xi\in\cH|\exists \psi\in\cH_1:
\xi=\psi\tens\eta\}\subset\cH$ and let $\{\psi_i\tens\eta\}_{i\in I}$
be an orthonormal basis of $\cH_1$. Then, $P(\psi_k\tens\eta|\cS_\eta)$
recovers $P(\psi_k\tens\eta)$ up to a normalization factor which is
the inverse of $\sum_{i\in I}|\rho(\psi_i\tens\eta)|^2$.

\subsection{Probability conservation}

Observe now that the split of the state space $\cH$ into the
components $\cH_1\tens\cH_{\bar{2}}$ in the standard geometry of
parallel spacelike hyperplanes is of a rather special nature. Firstly,
each of the tensor components separately has an inner product
and these are such that they are compatible with the inner product on
$\cH$ in the sense that
\[
 \langle \psi\tens\eta,\psi'\tens\eta'\rangle_{\cH}=
 \langle \psi,\psi'\rangle_{\cH_1}
 \langle \eta,\eta'\rangle_{\cH_{\bar{2}}} ,
\]
as guaranteed by axioms (T2b) and (T3b). Thus, in the first example we
may choose
$\psi$ to be normalized and $\{\eta_i\}_{i\in I}$ becomes an
orthonormal basis of $\cH_{\bar{2}}$.
Secondly, the induced map
$\tilde{\rho}:\cH_1\to\cH_2$ should be an isomorphism (again, we refer
to a discussion of this later). Thus, by axiom (T4b)
it must conserve the inner product. This implies in the example that
$\sum_{i\in
I}|\rho(\psi\tens\eta_i)|^2$ equals unity since it may be written as
$\sum_{i\in I} |\langle \iota_2(\eta_i),
\tilde{\rho}(\psi)\rangle_{\cH_2}|^2$. By similar
reasoning, the normalization factor in the third example equals unity.

The splitting of the boundary state space into a tensor product in the
way just described may serve as a global way of determining some
part of the measurement process as conditional on another one. This
includes automatic normalizations. The map $\tilde{\rho}$
may then be seen as describing an ``evolution''. Its compatibility
with the inner products (usually called \emph{unitarity}) leads to
what is known as ``conservation of probability'', ensuring
in this context the consistency of the interpretation. This is the
deeper meaning of axiom (T4b). Note that it only applies if
$\tilde{\rho}$ is invertible, otherwise it makes no sense to talk
about ``conservation''.
As shown in a concrete example in the companion paper \cite{Oe:KGtl},
such a splitting of the state space can also occur in cases where the
boundary does not decompose into disconnected components. Of course it
is then not axiomatically enforced, but part of the given theory.

The most general way of expressing ``probability conservation'' in the
present formalism (as ensured by axiom (T4b))
may be described as follows. Let $M$ and $N$ be
manifolds with disjoint interiors. Let $\Sigma$ be the boundary of the
union $M\cup
N$ and $\Sigma'$ be the boundary of $M$. Denote the associated state
spaces with $\cH$ and $\cH'$. Then $\rho_N$ gives rise to  a map
$\tilde{\rho}_N:\cH\to\cH'$.
Suppose that this map is invertible. Let
$\cA\subseteq\cS\subset\cH$ be closed subspaces. Denote their images
under $\tilde{\rho}_N$ by $\cA'\subseteq\cS'\subset\cH'$. Then, the
following equality of conditional probabilities holds,
\[
 P(\cA|\cS)=P(\cA'|\cS') .
\]

\begin{figure}
\begin{center}
\input{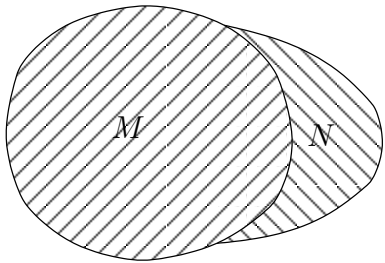}
\end{center}
\caption{A region $M$ with an adjacent smaller region $N$,
  ``deforming'' it.}
\label{fig:deform1}
\end{figure}

To provide an intuitive context of application for the above consider
the following. Let $M$ be some spacetime region. Now consider a
``small'' region $N$, adjacent to $M$ such that $M\cup N$ may be
considered a ``deformation'' of $M$, see Figure~\ref{fig:deform1} for
an illustration. Then, as described above, the
amplitude map for $N$ gives rise to the map $\tilde{\rho}_N$
interpolating between the state spaces associated with the boundary of
$M$ and its deformation $M\cup N$. Since we are dealing with a
``small'' deformation this map should be an isomorphism and
consequently preserve the inner product. Then, we may say that
probabilities are ``conserved under the deformation''.

\section{Vacuum axioms}
\label{sec:vacuum}

The main property of the vacuum state in standard quantum field theory
is its invariance under time-evolution. In the present context we
expect a family of vacuum states, namely one for each
oriented hypersurface. However, we will continue to talk about ``the''
vacuum state, since the members of this family should be related to
each other through certain coherence conditions. It is quite
straightforward to formulate these coherence conditions in axiomatic
form.

\begin{itemize}
\item[(V1)] For each hypersurface $\sig$ there is a distinguished state
  $\psi_{\sig,0}\in\cH_\sig$, called the \emph{vacuum state}.
\item[(V2)] The vacuum state is compatible with the involution. That is,
  for any hypersurface $\sig$,
  $\psi_{\osig,0}=\iota_\sig(\psi_{\sig,0})$.
\item[(V3)] The vacuum state is multiplicative. Suppose the hypersurface
  $\sig$ decomposes into disconnected components
  $\sig_1\cup\sig_2$. Then
  $\psi_{\sig,0}=\psi_{\sig_1,0}\tens\psi_{\sig_2,0}$.
\item[(V4)] The vacuum state is normalized. On any hypersurface $\Sigma$,
  $\langle\psi_{\sig,0},\psi_{\sig,0}\rangle=1$.
\item[(V5)] The amplitude of the vacuum state is
  unity, $\rho_M(\psi_{\partial M,0})=1$.
\end{itemize}

An important consequence of these properties in combination with the
core axioms is that they enforce ``conservation'' of the vacuum under
generalized
evolution, generalizing time-translation invariance.
Consider the situation of axiom (T4b). That is, we have a
region $M$ with boundary decomposing into disconnected components $\sig_1$
and $\sig_2$ and the amplitude gives rise to an isomorphism of vector
spaces $\tilde{\rho}_M:\cH_{\sig_1}\to\cH_{\osig_2}$. The image of
the vacuum state
$\psi_{\sig_1,0}$ under $\iota_{\osig_2}\circ\tilde{\rho}_M$ obviously
may be written as a linear combination
$\alpha \psi_{\sig_2,0}+\beta \psi_{\sig_2,1}$ where $\alpha$ and
$\beta$ are complex numbers and $\psi_{\sig_2,1}$ is a normalized
state orthogonal to the vacuum state $\psi_{\sig_2,0}$. On the other
hand, by definition of $\tilde{\rho}_M$ we have the equality
$\rho_M(\psi_{\sig_1,0}\tens\psi_{\sig_2,0})
=\langle \iota_{\osig_2}\circ\tilde{\rho}_M(\psi_{\sig_1,0}),
\psi_{\sig_2,0}\rangle_{\sig_2}$. Properties (V3)
and (V5) of the vacuum
then imply $\alpha=1$. On the other hand axiom (T4b) implies
preservation of the norm by $\tilde{\rho}_M$ and hence by
$\iota_{\osig_2}\circ\tilde{\rho}_M$. Since the vacuum is normalized
by axiom (V4) this forces $|\alpha|^2+|\beta|^2=1$. Hence, $\beta=0$.

Conversely, we may use this conservation property to transport the
vacuum state from one hypersurface to another one. This might lead one
to suggest the following prescription for the vacuum state
\cite{Ati:tqft,CDORT:vacuum}. Consider a region $M$ with boundary
$\sig$.
The amplitude $\rho_M:\cH_{\sig}\to\C$ gives rise to a linear
map $\tilde{\rho}_M:\C\to\cH_\osig$ by dualization and hence to a
state $\psi_M\in\cH_\osig$. In fact, this is ``almost''
true. Namely, in the general case of infinite dimensional state
spaces we should expect this state $\psi_M$ not
to be normalizable and hence not to exist in the strict sense. Let us
ignore this problem. It is easy to see that by its very definition
this state is automatically conserved via axiom (T5) in the way
described above. Nevertheless, it is not a good candidate for a vacuum
state in the sense of ``ground state'' or ``no-particle state''.
Namely, given such a vacuum state $\psi_{\osig,0}\in\cH_\osig$
there should be some ``excited state'' $\psi_s\in\cH_\osig$ that is
orthogonal to it and produces a non-zero amplitude via
$\rho_M(\iota_\osig(\psi_s))\neq 0$.
However, by construction of
$\psi_M$, we have $\langle \psi_s, \psi_M\rangle_\osig
= \rho_M(\iota_\osig(\psi_s))\neq 0$. Hence $\psi_M$
cannot be (a multiple of) the vacuum state $\psi_{\osig,0}$. Note
that to arrive at this conclusion we have used only the core axioms,
but none of the properties proposed above for the vacuum.

A single, uniquely defined vacuum state per hypersurface
represents merely the simplest possibility for realizing the concept
of a vacuum. A rather obvious generalization would be to a subspace of
``vacuum states'' per hypersurface. Approaches
to quantum field theory in curved space time indeed indicate that this
might be required \cite{Wal:qftcurved}. We limit ourselves here to the
remark that it is rather straightforward to adapt the above axioms to
such a context.

\section{Backgrounds and spacetime symmetries}
\label{sec:bgsym}

\subsection{Background structures}

The general boundary formulation is supposed
to be applicable to contexts where the basic spacetime objects
entering the formulation, namely regions and hypersurfaces, may have a
variety of meanings. In Section~\ref{sec:axioms} we already mentioned
the context corresponding to standard quantum field theory. Namely,
spacetime is Minkowski space, regions are 4-dimensional submanifolds
and hypersurfaces are closed oriented 3-dimensional submanifolds.

The context with a minimal amount of structure
is that of \emph{topological} manifolds with a given dimension $d$. Thus,
regions would be $d$+1-dimensional topological manifolds and
hypersurfaces would be closed oriented $d$-dimensional topological
manifolds. This is the context where the axioms are most closely
related to topological quantum field theory \cite{Ati:tqft}. An
additional layer of structure is given by considering
\emph{differentiable} manifolds, i.e., we add a
differentiable structure. Another layer of structure that is crucial
in ordinary quantum field theory is the (usually pseudo-Riemannian)
\emph{metric} structure. There are a variety of other structures of
potential interest in various contexts such as complex structure, spin
structure, volume form etc.

Any structure additional to the topological or differentiable one is
usually referred to as a \emph{background}. (Sometimes this terminology
includes the topological structure as well.) In addition to
considering the core axioms within different types of backgrounds
we can make a further choice. Namely, we might regard the regions and
hypersurfaces as manifolds in their own right, each equipped with its
prescribed background structure. Then, boundaries inherit the background
structure from the region they bound and the gluing of regions must
happen in such a way that the background structure is respected. On
the other hand, we might prescribe a global spacetime in which regions
and hypersurfaces appear as submanifolds of codimension 0 and 1
respectively. In this case, the spacetime manifold carries the
background structure which is inherited by the regions and
hypersurfaces. To distinguish the two situations we will refer to the
former as a \emph{local} background and to the latter as a
\emph{global} background. For example, in the standard quantum field
theoretic context we choose a global Minkowski background.

Let us briefly discuss various background structures appropriate in
a few situations of interest. As already mentioned, the natural
choice for standard quantum field theory is that of a
global Minkowski spacetime background. If we are interested in quantum
field theory on curved spacetime we might simply replace Minkowski
space with another global metric background spacetime. However, if we
wish to describe quantum field theory on curved spacetimes in
general, we might want to use \emph{local} metric
backgrounds. This would implement a \emph{locality} idea inherent in
the general boundary formulation, namely that processes happening in a
given region of spacetime are not dependent on the structure of
spacetime somewhere else.
In conformal field theory we would have $d=1$ and local
complex background structures. Indeed, Segal's axiomatization of
conformal field theory along such lines at the end of the
1980's \cite{Seg:cftproc,Seg:cftdef} had a seminal influence on the
mathematical framework
of topological quantum field theory as expressed in Atiya's
formalization \cite{Ati:tqft}.

Finally, in a hypothetical quantum theory of general relativity there
would be no metric background. Due to the background differential
structure inherent in classical general relativity one might expect
the same choice of background in the quantum theory, i.e., merely a
local differentiable structure. It is also conceivable that one has to
be more general and consider merely topological manifolds.
A relevant discussion can be found in \cite{Pfe:qgrclass}.
Even more exotic ``sums over topologies'' may be considered,
going back to a proposal of Wheeler
\cite{Whe:natgeometro}. Implementing these would require a
modification of the present framework.

\subsection{Symmetries}
\label{sec:sym}

Spacetime transformations act on regions and hypersurfaces.
It is natural to suppose that these induce algebraic transformations
on state spaces and amplitude functions.
In topological quantum field theory such transformations indeed
usually form an integral part of the framework \cite{Ati:tqft}.
On the other hand we are all familiar with the importance of the
Poincar\'e group and its representations for quantum field theory.

Spacetime transformations are intimately related to the background
structure. We may consider rather general
transformations (e.g., homeomorphisms or diffeomorphisms) or only such
transformations that leave a background structure
invariant. Furthermore, a crucial difference arises depending on
whether the background is global or local.
In the former case we consider transformations of the given spacetime
as a whole. These then induce transformations of or between regions
and hypersurfaces considered as
submanifolds. In the latter case we consider transformations of a
region or hypersurface considered as a manifold with background
structure in its own right. In particular, each region or hypersurface
a priori comes equipped with its own transformation group.

In standard quantum field theory we consider only transformations that
leave the global Minkowski background invariant. That is, the group of
spacetime transformations is the group of isometries of Minkowski
space, the Poincar\'e group. If we consider quantum field theory on
another global metric background we might equally restrict spacetime
transformations to isometries. More general transformations would make
sense if we wish to consider an ensemble of
backgrounds. Alternatively, if we are interested in quantum field
theory in general curved spacetime utilizing local backgrounds we
would use general transformations, too, probably
diffeomorphisms. But these would be \emph{local} diffeomorphisms of
the regions and hypersurfaces themselves and not \emph{global} ones, of a
whole spacetime. The latter transformations seems also the most
natural ones for a quantum theory without metric background (such as
quantum general relativity). Of course, in that case there is no metric
background which they modify.

Since the natural transformation properties of state spaces and
amplitude functions take a somewhat different form depending on
whether we are dealing with a global or a local background we will
separate the two cases. We start with the case of a global
background.

\subsubsection{Global backgrounds}

Let $G$ be a group of transformations acting on spacetime. We demand
that this group maps regions to regions and hypersurfaces to
hypersurfaces. (Recall that there are generally restrictions as to
what $d$+1-submanifold qualifies as a region and what closed oriented
d-manifold qualifies as a hypersurface, see
Section~\ref{sec:corners}.) Let $g\in
G$. We denote the image of a hypersurface $\sig$ under $g$ by
$g\act\sig$. Similarly we denote the image of a region $M$ under $g$
as $g\act M$. We postulate the following axioms.

\begin{itemize}
\item[(Sg1)] The
  action of $G$ on hypersurfaces induces an action on the ensemble of
  associated state spaces. That is, $g\in G$
  induces a linear isomorphism
  $\cH_\Sigma\to\cH_{g\act\Sigma}$, which we denote on
  elements as $\psi\mapsto g\act\psi$. It has the properties of a
  (generalized) action, i.e.,
  $g\act(h\act \psi)=(g h)\act\psi$ and
  $e\act\psi=\psi$, where $e$ is the identity of $G$.\footnote{Note
  that in spite of the suggestive
  notation this is \emph{not} an action in the usual sense. Indeed, a
  group element here generally maps a state from one space to a state
  in a different state space. Nevertheless we will use the word
  ``action'' for simplicity.}
\item[(Sg2)] The action of $G$ on state spaces is compatible with the
  involution. That is,
  $\iota_{g\act\sig}(g\act\psi)=g\act\iota_\sig(\psi)$ for
  any $g\in G$ and any hypersurface $\sig$.
\item[(Sg3)] The action of $G$ on state spaces is compatible with the
  decomposition of hypersurfaces into disconnected components. Suppose
  $\sig= \sig_1\cup\sig_2$ is such a decomposition, then we require
  $g\act(\psi_1\tens\psi_2)=g\act\psi_1\tens g\act\psi_2$ for any
  $g\in G$, $\psi_1\in \cH_{\sig_1}$, $\psi_2\in\cH_{\sig_2}$.
\item[(Sg4)] The action of $G$ on state spaces is compatible with the
  bilinear form. That is, $(g\act\psi_1,g\act\psi_2)_{g\act\sig}
 =(\psi_1,\psi_2)_\sig$.
\item[(Sg5)] The action of $G$ on regions leave the amplitudes
  invariant, i.e., $\rho_{g\act M}(g\act\psi)=\rho_M(\psi)$ where $M$
  is any region, $\psi$ any vector in the state space associated to
  its boundary.
\item[(SgV)] The vacuum state is invariant under $G$, i.e.,
  $g\act\psi_{\sig,0}=\psi_{g\act\sig,0}$.
\end{itemize}

\subsubsection{Local backgrounds}

We now turn to the case of local backgrounds. In this case we
associate with each region $M$ its own transformation group $G_M$ that
maps $M$ to itself (but with possibly modified background). In
particular, $G_M$ preserves boundaries. Similarly,
each hypersurface $\sig$ carries its own transformation group
$G_\sig$, mapping $\sig$ to itself (again with possibly modified
background). Furthermore, we demand that for any region $M$ with
boundary $\sig$ there is a group homomorphism $G_M\to G_\sig$ that
describes the induced action of $G_M$ on the boundary. We denote the
image of $\sig$ under $g\in G_\sig$ by $g\act\sig$.
Similarly, we denote the image of the
region $M$ under the action of $G_M$ by $g\act M$. We use the same
notation for the induced action on the boundary $\sig$ of $M$.

\begin{itemize}
\item[(Sl1)] The action of $G_\sig$ on $\sig$ induces an action on the
  ensemble of state spaces associated with the different background
  structures of $\sig$. That is, $g\in G_\sig$
  induces a linear isomorphism
  $\cH_\Sigma\to\cH_{g\act\Sigma}$, which we denote on
  elements as $\psi\mapsto g\act\psi$. It has the properties of a
  (generalized) action, i.e.,
  $g\act(h\act \psi)=(g h)\act\psi$ and
  $e\act\psi=\psi$, where $e$ is the identity of $G_\sig$.
\item[(Sl2)] $G_\sig$ is compatible with
  the involution. That is, $G_\osig=G_\sig$ are canonically identified,
  with $\iota_{g\act\sig}(g\act\psi)=g\act\iota_\sig(\psi)$ for
  any $g\in G_\sig$ and any hypersurface $\sig$.
\item[(Sl3)] $G_\sig$ is compatible with the decomposition of
  hypersurfaces into disconnected components. Suppose
  $\sig= \sig_1\cup\sig_2$ is such a decomposition. Consider the
  subgroup $G_\sig'\subseteq G_\sig$ that maps the components to
  themselves. Then,
  $G_\sig'=G_{\sig_1}\times G_{\sig_2}$ such that
  $(g_1,g_2)\act(\psi_1\tens\psi_2)=g_1\act\psi_1\tens g_2\act\psi_2$
  for any
  $g_1\in G_{\sig_1}$, $g_2\in G_{\sig_2}$, $\psi_1\in \cH_{\sig_1}$,
  $\psi_2\in\cH_{\sig_2}$.
\item[(Sl4)] $G_\sig$ is compatible with the
  bilinear form. That is, $(g\act\psi_1,g\act\psi_2)_{g\act\sig}
 =(\psi_1,\psi_2)_\sig$.
\item[(Sl5)] $G_M$ leaves the amplitude $\rho_M $
  invariant, i.e., $\rho_{g\act M}(g\act\psi)=\rho_M(\psi)$ where $M$
  is any region, $\psi$ any vector in the state space associated to
  its boundary.
\item[(SlV)] The vacuum state is invariant under $G_\sig$, i.e.,
  $g\act\psi_{\sig,0}=\psi_{g\act\sig,0}$.
\end{itemize}

These axioms, both in the global as well as in the local case are
supposed to describe only the most simple situation. It might be
necessary to modify them, for example introducing phases, cocycles
etc.

\subsection{Invertible evolution}
\label{sec:invevol}

Let us return to a question that has arisen in
Section~\ref{sec:recover} in the context of recovering the standard
formulation of quantum mechanics. Consider a time interval $[t_1,t_2]$
giving rise to a corresponding region $M$ of spacetime. Denote
the two components of the bounding hypersurface by $\sig_1$ and
$\osig_2$ respectively, see Figure~\ref{fig:standard}. Firstly, the
core axioms do not tell us that
there is a (natural) isomorphism between the state spaces
$\cH_{\sig_1}$ and $\cH_{\sig_2}$. However, it is clear that this is
related to time translations. Indeed, we are in the
context of a global metric background and suppose that its isometry
group $G$ includes time translations. A time translation $g_\Delta\in
G$ by the amount
$\Delta=t_2-t_1$ maps $\sig_1$ to $\sig_2$. Thus, by axiom (Sg1) the
state spaces $\cH_{\sig_1}$ and $\cH_{\sig_2}$ are identified through
the induced action. Indeed, we may use time translations to
identify all equal-time hypersurfaces in this way, arriving at
\emph{the} state space of quantum mechanics.

A second point noted in Section~\ref{sec:recover} is that even given
natural isomorphisms between the state spaces, it does not follow from
the core axioms that the amplitude function
$\rho_M:\cH_{\sig_1}\tens\cH_{\sig_2}\to\C$ yields an isomorphism of vector
spaces $\tilde{\rho}_M:\cH_{\sig_1}\to\cH_{\sig_2}$.
It should also be clear why we cannot simply enforce this on the level
of the core axioms. Namely, intuitively, we only expect an isomorphism
if $M$ connects in a suitable way $\sig_1$ and $\sig_2$ and if
$\sig_1$ and $\sig_2$ have the ``same size''. We will come back to the
discussion of ``sizes'' of state spaces in Section~\ref{sec:size}.

Enforcing the existence of an isomorphism $\tilde{\rho}_M$ in
suitable situations may be achieved along the lines of the following
procedure using an isotopy.
Let $G$ by the transformation group of the global background $B$ in
question. Suppose there is a smooth map $\alpha:I\to G$ from the unit
interval to $G$ such that $\alpha(0)=e$, ($e$ the neutral element of
$G$) and $\alpha(1)\act\sig_1=\sig_2$. Furthermore, assume that the
induced map $I\times\sig\to B$ has image $M$ and is a
diffeomorphism (or just homeomorphism in the absence of differentiable
structure) onto its image. Then, require that the amplitude map
induces an isomorphism of vector spaces (and by axiom (T4b) thus of
Hilbert spaces) $\tilde{\rho}_M:\cH_{\sig_1}\to\cH_{\osig_2}$. This
prescription would apply
in particular to the standard formulation, enforcing an invertible (and
by axiom (T4b) thus unitary) time-evolution operator as required.

\section{Schr\"odinger representation and Feynman integral}
\label{sec:schrfeyn}

The Schr\"odinger representation, i.e., the representation of states
in terms of wave functions, together with the Feynman path integral
provide a natural context for the realization of the general boundary
formulation \cite{Oe:boundary,CDORT:vacuum}. The former facilitates an
intuitive
implementation of the axioms relating to states, while the latter
(seems to) automatically satisfy the composition axiom (T5). Let us
give a rough sketch of this approach in the following.

We suppose that there is a \emph{configuration} space $K_\sig$
associated to every hypersurface $\sig$. We define the
state space $\cH_\sig$ to be the space of (suitable) complex valued
functions on
$K_\sig$, called \emph{wave functions}, providing (T1). (This was
denoted (Q1) in \cite{Oe:boundary}.)
We suppose that $K_\sig$
is independent of the orientation of $\sig$. Thus, the state spaces on
$\sig$ and its oppositely oriented version $\osig$ are the same, 
$\cH_\sig=\cH_\osig$. The antilinear involution
$\iota_\sig:\cH_\sig\to\cH_\osig$ is given by the complex conjugation
of functions. That is, for any wave function $\psi\in\cH_\sig$ and any
configuration $\varphi\in K_\sig$ we have
$(\iota_\sig(\psi))(\varphi)=\overline{\psi(\varphi)}$, satisfying
(T1b).

We suppose that the configuration space on a hypersurface $\sig$
consisting of disconnected components $\sig_1$ and $\sig_2$ is the
product of the individual configuration spaces,
$K_\sig=K_{\sig_1}\times K_{\sig_2}$. This implies,
$\cH_\sig=\cH_{\sig_1}\tens\cH_{\sig_2}$, i.e., (T2). Since the
complex conjugate of a product is the product of the complex
conjugates of the individual terms (T2b) follows.

Given a measure on the configuration spaces, a bilinear form
$\cH_\osig\tens\cH_\sig\to\C$ is defined via
\begin{equation}
 (\psi,\psi')_\sig=\int_{K_\sig}\xD\varphi\,
  \psi(\varphi)\psi'(\varphi) .
\end{equation}
This induces the inner product
\begin{equation}
  \langle\psi,\psi'\rangle_\sig=\int_{K_\sig}\xD\varphi\,
  \overline{\psi(\varphi)}\psi'(\varphi) .
\label{eq:schrip}
\end{equation}
This yields (T3). Since the integral over a product of spaces is the
product of the integrals over the individual spaces we have (T3b).

Let $M$ be a region with boundary $\sig$. The amplitude of a wave
function $\psi\in\cH_\sig$ is given by the following heuristic path
integral formula, providing (T4),
\begin{equation}
 \rho_M(\psi)=\int_{K_\sig}\xD\varphi\, \psi(\varphi) Z_M(\varphi),
 \quad\text{with}\quad
 Z_M(\varphi)=\int_{K_M, \phi|_\sig=\varphi}\xD\phi\,
 e^{\frac{\im}{\hbar} S_M(\phi)} .
\label{eq:pintampl}
\end{equation}
(This was denoted (Q2) in \cite{Oe:boundary}.)
The second integral is over ``all field configurations'' $\phi$ in the
region $M$ that reduce to $\varphi$ on the boundary. $S_M$ is the
action integral over the region $M$. The quantity $Z_M(\varphi)$ is also
called the \emph{field propagator}. It formally looks like a wave
function and thus like a state. Indeed, this is precisely the state
$\psi_M$ briefly discussed at the end of Section~\ref{sec:vacuum}. As
already mentioned there, $\psi_M$ is in general not normalizable
and thus not a state in the strict sense.\footnote{In
\cite{CDORT:vacuum} the field propagator
$Z$ was denoted by $W$. Furthermore, the word ``vacuum''
was used for the state $\psi_M$. However, as explained at the end of
Section~\ref{sec:vacuum} such a state has nothing to do with the more
usual notion of vacuum considered there. We thus discourage
the use of the term ``vacuum'' for this state.}

Consider a region $M$ with boundary $\sig$ decomposing into two
disconnected components $\sig_1$ and $\sig_2$. Then, we can
immediately write down the formal map $\tilde{\rho}_M:\cH_{\sig_1}
\to\cH_{\osig_2}$ induced by the amplitude map $\rho_M$. Namely,
\begin{equation}
 (\tilde{\rho}_M(\psi))(\varphi')
 =\int_{K_{\sig_1}}\xD\varphi\, \psi(\varphi) Z_M(\varphi,\varphi'),
\end{equation}
where $\varphi'\in K_{\osig_2}$. Of course, the strict existence of
the resulting state depends on its normalizability. Suppose that it
does exist and that $\tilde{\rho}_M$ provides an isomorphism of
vector spaces. It is then easy to
see that the validity of axiom (T4b), i.e., the preservation of the
inner product (\ref{eq:schrip}) or \emph{unitarity} would follow from
the formal equality
\begin{equation}
 \int_{K_{\sig_2}}\xD\varphi_2\, \overline{Z_M(\varphi_1,\varphi_2)}
 Z_M(\varphi_1',\varphi_2)=\delta(\varphi_1,\varphi_1'),
\end{equation}
for $\varphi_1,\varphi_1'\in K_{\sig_1}$. This basically says that
the conjugate of the propagator describes the inverse of the
original propagator.

Finally, in the context of axiom (T5) the condition
$\tilde{\rho}_{M_1\cup M_2}=\tilde{\rho}_{M_2}\circ\tilde{\rho}_{M_1}$
translates to the following condition on propagators,
\begin{equation}
 \int_{K_{\osig}}\xD\varphi\, Z_{M_1}(\varphi_1,\varphi)
  Z_{M_2}(\varphi,\varphi_2) = Z_{M_1\cup M_2}(\varphi_1,\varphi_2) ,
\label{eq:pintcomp}
\end{equation}
with $\varphi_1\in K_{\sig_1}$ and $\varphi_2\in K_{\osig_2}$.
If we write the propagator in terms of the path integral
(\ref{eq:pintampl}) the validity of (\ref{eq:pintcomp}) becomes
obvious. Namely, it just says that we may choose a slice in a region
and split a path integral over the region as follows:
One integral over configurations in the slice and an integral over
the whole region restricted to configurations matching the given one
on the slice. Thus, (T5) holds.

In fact, the picture presented so far, while being rather compelling,
turns out to be somewhat too naive. For example, it was shown in
\cite{Oe:timelike} (in the context
of the Klein-Gordon theory) that the configuration space on a timelike
hyperplane is not simply the space of ``all'' field
configurations, but a smaller space of \emph{physical} configurations.
Remarkably, the composition rule (\ref{eq:pintcomp}) works with this
restricted configuration space on the intermediate slice rather than
the ``full'' configurations space one would obtain by naively slicing
the spacetime path integral. Other non-trivial issues include
normalization factors, which as one might expect turn out to be
generically infinite. Nevertheless, \cite{Oe:timelike} and even more
so the companion article \cite{Oe:KGtl} show (for the Klein-Gordon
theory) that the Schr\"odinger-Feynman approach to the implementation
of the general boundary formulation is a viable one.

\section{The shape of regions and the size of state spaces}
\label{sec:size}

\begin{figure}
\begin{center}
\input{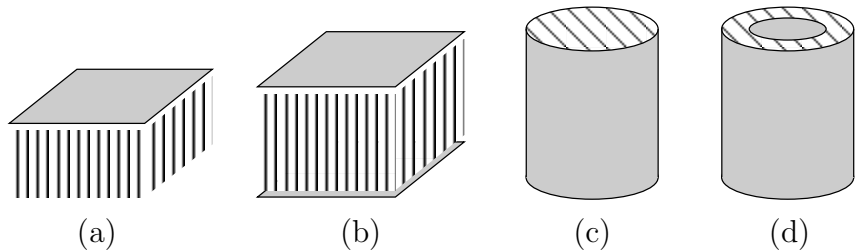}
\end{center}
\caption{Examples of admissible and inadmissible regions
  (regions are hatched, hypersurfaces shaded): Inadmissible: (a) the
  half-space. Admissible: (b) parallel hyperplanes, (c) solid
  hypercylinder, (d) nested hypercylinders.}
\label{fig:regions}
\end{figure}

So far we have been rather vague about what kind of regions and what
kind of hypersurfaces are actually admissible. For simplicity, let us
discuss this question in the context of a global Minkowski
background. It is then clear that regions are 4-dimensional
submanifolds and hypersurfaces are 3-dimensional closed oriented
submanifolds. In fact, given that we know which regions are
admissible, we can easily say which hypersurfaces are
admissible. Namely, any hypersurface is admissible that arises as the
boundary of an admissible region or a connected component thereof. So,
which regions are admissible?

Unfortunately, at this stage we do not have a complete answer to this
question. In the following we give some partial answers that are
mainly obtained through experience with the application of this
framework to quantum field theory in general (along the lines of
Section~\ref{sec:schrfeyn}) and
the Klein-Gordon theory \cite{Oe:timelike,Oe:KGtl} in particular. As
should be clear from Section~\ref{sec:prob}, a region must
be such that we can associate with it a ``complete measurement
process''. In terms of the standard formulation this means preparation
plus observation. Thus, in that formulation the type of region of main
interest is a time interval times all of space
(Figure~\ref{fig:regions}.b).

On the other hand, consider a region consisting of the past (or
future) half of Minkowski space. In this situation the standard
formulation applies as well. Indeed, it tells us that no complete
measurement process can be associated with just one equal-time
hyperplane (Figure~\ref{fig:regions}.a). This gives us two examples: A
time-interval
defines an admissible region while a (temporal) half-space does not
define an admissible region. How can this be generalized?

It turns out that a useful way to think about admissible regions is
that the configuration data on the boundary is essentially in
one-to-one correspondence to classical solutions. (Recall that we use
the context of Section~\ref{sec:schrfeyn}.) Indeed, this
correspondence is used in \cite{Oe:boundary, Oe:KGtl} to
calculate the field propagator (\ref{eq:pintampl}).
This qualifies the
time-interval as admissible (knowing the field configuration at two
times essentially determines a solution) while it disqualifies the
half-space as inadmissible (there are many solutions restricting to
the same field configuration at a given time).

In \cite{Oe:boundary} the explicit examples of admissible regions
were extended to regions enclosed between any two parallel
hyperplanes (spacelike or timelike). In the companion paper
\cite{Oe:KGtl} further examples are considered, in particular, a full
hypercylinder. More precisely, this is a ball in space times the time
axis. Again, its boundary data is in correspondence to classical
solutions.
Note that such a situation is impossible in a traditional context of
only spacelike hypersurfaces.
We have here an example where a \emph{connected}
boundary carries states describing a complete measurement process.
In particular, this implies that there is no a priori distinction
between the ``prepared'' and the ``observed'' part of the measurement
process. Hence it goes beyond the applicability of 
the standard probability interpretation, highlighting the necessity
for and meaning of a generalized interpretation as outlined in
Section~\ref{sec:prob}.

Since the configuration data on the hypercylinder as well as on two
parallel hyperplanes correspond to classical solutions one might say
that the associated state spaces have ``equal size''. Let us
call these state spaces of size $1$. A single hyperplane carries half
the data and we say the associated state
space has size $\frac{1}{2}$. In this way the size of state spaces is
additive with respect to the disjoint union of the underlying
hypersurfaces.

Another valid way to obtain admissible regions should be by
forming a disjoint union of admissible regions. Physically, that means
that we are performing several concurrent and independent
measurements. (The word ``concurrent'' should be understood here in a
logical rather than a temporal sense here.) We can reconcile this with
the idea of a correspondence between boundary data and a classical
solution if we restrict the solution to the region itself, rather than
it being defined globally.
Indeed, this intuition receives independent confirmation from the
second new example of \cite{Oe:KGtl}. This is a region formed by a thick
spherical shell in space times the time axis. Its boundary consists of
two concentric hypercylinders. It turns out that the data on the boundary
is in correspondence to classical solutions defined on the region, but
generically containing singularities outside. Furthermore, following
this principle of correspondence yields the correct field propagator
consistent with the other results.
In terms of the terminology introduced above, the
state space associated with the boundary of this region has size
$2$.

It thus appears that a region should be admissible if the
configuration data on its boundary is essentially in correspondence to
classical solutions defined inside the region. Note also that the size
of the boundary state space of an admissible region seems to be
necessarily an integer. Of course we expect the heuristic arguments
put forward here, even if based on limited examples, to be
substantially modified or generalized in a fully worked out theory.

Finally, our discussion was largely oriented at quantum field
theory with metric backgrounds.
It \cite{Oe:catandclock} it was argued (motivated by the problem of
time) that in the context of a quantum theory of spacetime, valid
measurements should correspond to regions with a connected boundary
only. Thus, we might expect such a further limitation on the
admissibility of regions in that context.

\section{Corners and empty regions}
\label{sec:corners}

There is another aspect concerning the shape of regions that merits
attention and points to some remaining deficiencies of the treatment
so far. Most experiments are constrained in space and time and so it
seems most natural to describe them using finite regions of
spacetime. Indeed, as it was argued in \cite{Oe:catandclock}, the
region of generic interest is that of a topological 4-ball.

The most elementary composition we might think of in this context is
that of two 4-balls to another 4-ball. However, this
raises an immediate problem: To merge two 4-balls into one, we would
need to glue them at parts of their boundaries only. Thus, we would
need to distinguish between different parts of a connected
hypersurface and glue only some of them. This is clearly not covered
by the composition axiom (T5) as it stands.

If the regions have not only topological, but also differentiable
structure (as in almost every theory of conceivable physical interest)
the problem is even more serious. Namely, we need to allow
\emph{corners} in the boundaries of regions to glue them
consistently. One way to think about corners is as ``boundaries of
boundaries''. Somewhat more precisely, the places on the boundary
where the normal vector changes its direction discontinuously, are the
corners. A simple example of a region with corners is a 4-cube (which
is topologically a 4-ball of course). It
has the important property that gluing two 4-cubes in the obvious way
yields another 4-cube shaped region.

\begin{figure}
\begin{center}
\input{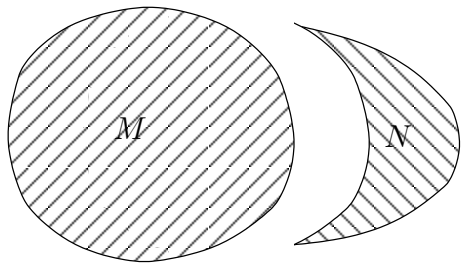}
\end{center}
\caption{The region $N$ of Figure~\ref{fig:deform1} has corners.}
\label{fig:deform2}
\end{figure}

We now recognize that the region $N$ in Figure~\ref{fig:deform1}
actually contains
corners and thus, strictly speaking, falls outside of axiom (T5).
See Figure~\ref{fig:deform2}. In
any case, what we need is a further extension of the core axioms to
accommodate corners. Probably, we need to allow a splitting of state
spaces along corners. This could be subtle, though, possibly involving
extra data on the corner relating the two state spaces etc. In
topological quantum field theory, the subject of corners already has
received some attention, see e.g.\ \cite{Wal:tqftnotes}. However, at
this point we will not speculate on how they may be implemented into
the present framework.

It is interesting to note that in some situations corners can be
avoided by a different generalization of regions which is rather
natural in our framework. This generalization consists in allowing
regions to consist partly or even entirely purely of
boundaries. These regions are partly or entirely
``empty''. Note that in this context an empty part should be thought
of as having \emph{two} distinct (and disconnected) boundary
hypersurfaces, namely
one for each orientation. A ``completely empty'' region is
determined simply by one given hypersurface $\sig$. The ``boundary''
of this region is the hypersurface $\sig$ together with its opposite
$\osig$, to be thought of as disconnected in the sense of axiom (T2).

\begin{figure}
\begin{center}
\input{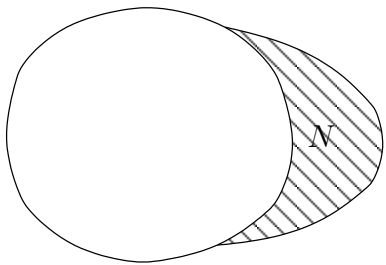}
\end{center}
\caption{An alternative version of the region $N$ with ``empty''
  parts, but without corners.}
\label{fig:deform3}
\end{figure}

Indeed, reconsider the example of Figure~\ref{fig:deform1}. If we
allow $N$ to be
partly empty, we can avoid the need for corners. This is shown in
Figure~\ref{fig:deform3}. Indeed, recall the example at the end of
Section~\ref{sec:prob}. The map $\tilde{\rho}_N$ is the map induced by
the amplitude map of $N$, extended by an identity. At least this is
the case in the context with corners. In the context where $N$ is the
region shown in Figure~\ref{fig:deform3} the map $\tilde{\rho}_N$ is
simply induced by the amplitude, without any extension.

The extension to ``empty'' regions does not require any changes in the
axioms. To the contrary, it actually simplifies some axioms and makes
their meaning more transparent. Of the core axioms, the main example
is (T3). The bilinear form postulated for a hypersurface $\sig$ is
nothing but the amplitude map for the empty region defined by
$\sig$. The symmetry of this bilinear form is then automatic (since
$\osig$ defines the same empty region). Thus,
axiom (T3) becomes almost entirely redundant, except for the
requirement that it induces (together with the involution) a Hilbert
space structure. This is simply a suitable non-degeneracy
condition. This also explains why we have formulated (T3) in such a way
that the bilinear form is fundamental and the inner product derived,
rather than the other way round.

Axiom (T3b) is then also redundant as it arises as a special case of
axiom (T5) when the intermediate hypersurface $\sig$ is
empty. Furthermore, axiom (T4b) is automatic for the completely empty
regions, being guaranteed by what remains of axiom (T3).

Not only the core axioms are simplified. The vacuum normalization
axiom (V4) becomes an automatic consequence of the unit amplitude
axiom (V5). Similarly, the symmetry axiom (Sg4) now follows from (Sg5)
with (Sg3) and (Sl4) follows from (Sl5) with (Sl3).

\section{What about operators?}
\label{sec:ops}

We seem to have avoided so far a subject of some prominence in quantum
mechanics: operators. This has several reasons. Firstly, the dynamics
of a quantum theory may be entirely expressed in terms of its transition
amplitudes. Indeed, in quantum field theory it is (an idealization of)
these which yield the S-matrix and hence the experimental predictions
in terms of scattering cross sections. Secondly, all principal topics
discussed so far (probability interpretation, vacuum, spacetime
symmetries etc.) can indeed be formulated purely in a state/amplitude
language. Thirdly, since there are now many state spaces, there are
also many operator spaces. What is more, an operator in the standard
picture might correspond to something that is not an operator in the
present formulation.

There is one (type of) operator that we actually have
discussed: The time-evolution operator. Indeed, in the present
formulation it is most naturally expressed as a function rather than
an operator. Note that we can do a similar reformulation with any
operator of the standard formalism. It may be expressed as a function
on the standard state space times its dual. This in turn might be
identified (via a time translation symmetry) with the total state
space of a time interval region. However, for a general operator the
resulting function might not be of particular physical relevance.

Consider for
example creation and annihilation operators in a Fock 
space context. It appears much more natural to have such operators
also on tensor product spaces rather then turning them into
functions there. In the example of the Klein-Gordon theory
\cite{Oe:timelike,Oe:KGtl}
such operators can indeed be constructed on state spaces of size
larger than $\frac{1}{2}$. (This will be shown elsewhere.)

To describe the probability interpretation (Section~\ref{sec:prob}) we
might as well have used orthogonal projection operators instead of
subsets. Indeed, using projection operators is more useful in
expressing consecutive measurements. Note that the word
``consecutive'' here is not restricted to a temporal context alone.
Indeed, we may sandwich projection operators (or any operators for
that matter) between regions by inserting them into the composition of
induced maps described in axiom (T5).

Note also that it makes perfect sense to talk about expectation values
of operators in a given state. Namely, let $O$ be an operator on the
Hilbert space $\cH_\sig$ associated with some hypersurface $\sig$,
then its expectation value with respect to a state $\psi\in\cH_\sig$
may be defined in the obvious way,
\[
 \langle O\rangle \defeq \langle \psi, O\psi\rangle_\sig .
\]
Clearly, this reduces to the standard definition in the standard
circumstances.

Let us make some general remarks concerning
operators in relation to the axioms. The involution of axiom (T1b)
induces for any hypersurface $\sig$
a canonical isomorphism between operators on the space
$\cH_\sig$ and operators on the space $\cH_\osig$ via $O\mapsto
\iota\circ O\circ\iota$. In the context of axiom (T4b) the amplitude
provides an isomorphism between operator spaces induced by the
isomorphism of state spaces. Spacetime symmetries acting on state
spaces induce actions on operator spaces in the obvious way.

Finally, we come back to the time evolution operator. Its
infinitesimal form, the Hamiltonian, plays a rather important role in
the standard formulation. Obviously, it is of much less importance here,
as it is related to a rather particular 1-parameter deformation of
particular hypersurfaces. Attempts have been made already in the
1940's to find a generalized ``Hamiltonian'' related to local
infinitesimal deformations of spacelike hypersurfaces
\cite{Tom:relwave,Sch:qed1}. More
recently, steps have been taken to generalize this to the
general boundary formulation
\cite{CDORT:vacuum,CoRo:genschroed,Dop:tomschwing}.

\section{Conclusions and Outlook}
\label{sec:concl}

We hope to have presented in this work a compelling picture which puts
the idea of a general boundary formulation of quantum mechanics on a
solid foundation. In
particular, we hope to have shown convincingly, how the probability
interpretation of standard quantum mechanics extends in a consistent
way, including generalizations of the notions of probability
conservation and unitarity. A concrete example of its application in a
situation outside of the range of applicability the standard
formulation can be found in the companion paper \cite{Oe:KGtl}.

Nevertheless, we wish to emphasize that the present proposal is still
tentative and should not be regarded as definitive. An obvious
remaining deficiency was elaborated on in
Section~\ref{sec:corners}. This is the need for corners of regions and
hypersurfaces. This will certainly require a further refinement of
the core axioms.

Another open issue of significant importance is that of
quantization. Although we have outlined in Section~\ref{sec:schrfeyn}
how a combined Schr\"odinger-Feynman approach provides an ansatz here,
it is clearly incomplete. In particular, one would like to have a
generalization of canonical quantization to the present framework. A
difficulty is the lack of a simple parametrizability of ``evolution'',
making an infinitesimal approach through a (generalized) Hamiltonian
difficult. Perhaps the ``local'' Hamiltonian approach mentioned at the
end of Section~\ref{sec:ops} can help here, although it is not clear
that it would not be plagued by ordering ambiguities.

As mentioned in the introduction, a main motivation for the general
boundary formulation has been its potential ability of rendering the
problem of quantization of gravity more accessible. Indeed, in the
context of a mere differentiable background, one may think of it as
providing a ``general relativistic'' version of quantum mechanics (or
rather quantum field theory). Steps to apply (some form) of this
framework to quantum gravity have indeed been taken, notably in the
context of the loop approach to quantum gravity
\cite{CDORT:vacuum,Oe:bqgrav,MoRo:scatlqg,Rov:gravprop}.
The general boundary idea has also been advocated by Rovelli in
his excellent book on loop quantum gravity
\cite{Rov:qg}. In any case, this direction of research is still at its
beginning, but we hope to have set it on a more solid foundation.

Of course, the general boundary formulation might be useful in other
approaches to quantum gravity as well, such as string theory. Indeed,
a hope could be that the possibility to define local amplitudes would
remove the necessity to rely exclusively on the asymptotic S-matrix with
its well known limitations (e.g, problems with de~Sitter spacetime
etc.). Of course, the technical task of implementing this might be
rather challenging.

We close by pointing out a possible conceptual relation to 't~Hooft's
holographic principle \cite{tHo:holographic}. As
mentioned in Section~\ref{sec:axioms}, an intuitive way to think about
states in a quantum process is as encoding the possibly available data
or information about the process. We might also say that the
states encode the ``degrees of freedom'' of the process. This is
reminiscent of the holographic principle, albeit in one dimension
higher. Here the
degrees of freedom of a 4-volume sit on its boundary hyperarea. What
is lacking at this level is of course a numerical relation between
the number of degrees of freedom and (hyper)area. However, one might
imagine that the relevant state space of a quantum theory of gravity
can be graded by hyperarea. The holographic principle would then be a
statement about the distribution of the eigenvalues of the ``hyperarea
operator''.

\bibliography{stdrefs}
\bibliographystyle{amsordx}

\end{document}